\documentclass[5p,times]{elsarticle}

\usepackage{hyperref}
\usepackage{amssymb}
\usepackage{subfigure}

\journal{Physics Letters B}

\begin{document}

\begin{frontmatter}

\title{Single-particle shell strengths near the doubly magic nucleus $^{56}$Ni 
and the $^{56}$Ni($p$,$\gamma$)$^{57}$Cu reaction rate in explosive 
astrophysical burning}

\author[edinburghaddy]{D.~Kahl\corref{mycorrespondingauthor}}
\ead{daid.kahl@ed.ac.uk}
\cortext[mycorrespondingauthor]{Corresponding author}
 
\author[edinburghaddy]{P.~J.~Woods}
\author[nscladdy,msuaddy,jinaaddy]{T.~Poxon-Pearson}
\author[nscladdy,msuaddy,jinaaddy]{F.~M.~Nunes}
\author[nscladdy,msuaddy,jinaaddy]{B.~A.~Brown}
\author[nscladdy,msuaddy,jinaaddy]{H.~Schatz}
\author[nscladdy]{T.~Baumann}
\author[nscladdy]{D.~Bazin}
\author[nscladdy]{J.~A.~Belarge}

\author[nscladdy]{P.~C.~Bender\corref{lowellalt}}
\fntext[lowellalt]{Present Address: Department of Physics and Applied Physics, University 
of Massachusetts Lowell, Lowell, MA 01854, USA.}

\author[nscladdy,msuaddy]{B.~Elman}
\author[cmuaddy]{A.~Estrade}
\author[nscladdy,msuaddy,jinaaddy]{A.~Gade}
\author[jyvaddy]{A.~Kankainen}
\author[edinburghaddy]{C.~Lederer-Woods}
\author[nscladdy,msuaddy,jinaaddy]{S.~Lipschutz}
\author[nscladdy,msuaddy]{B.~Longfellow}
\author[edinburghaddy]{S.-J.~Lonsdale}
\author[nscladdy,msuaddy]{E.~Lunderberg}
\author[nscladdy,jinaaddy]{F.~Montes}
\author[nscladdy,msuaddy,jinaaddy]{W.~J.~Ong}
\author[cmuaddy]{G.~Perdikakis}
\author[nscladdy,jinaaddy]{J.~Pereira}
\author[nscladdy,msuaddy,jinaaddy]{C.~Sullivan}
\author[nscladdy,msuaddy,jinaaddy]{R.~Taverner}
\author[nscladdy]{D.~Weisshaar}
\author[nscladdy,msuaddy,jinaaddy]{R.~Zegers}

\address[edinburghaddy]{School of Physics \& Astronomy, University of Edinburgh, Edinburgh EH9 3FD, United Kingdom}
\address[nscladdy]{National Superconducting Cyclotron Laboratory, Michigan State 
University, East Lansing, MI 48824, USA}
\address[msuaddy]{Department of Physics \& Astronomy, Michigan State University, 
East Lansing, MI 48824, USA}
\address[jinaaddy]{JINA Center for the Evolution of the Elements, Michigan State 
University, East Lansing, MI 48824, USA}
\address[cmuaddy]{Central Michigan University, Mount Pleasant, MI 48859, USA}
\address[jyvaddy]{University of Jyv\"askyl\"a, P.O. Box 35, FI-40014 University of 
Jyv\"askyl\"a, Finland}

\begin{abstract}

Angle-integrated cross-section measurements of the $^{56}$Ni($d$,$n$) and 
($d$,$p$) stripping reactions have been performed to determine the 
single-particle strengths of low-lying excited states in the mirror nuclei pair 
$^{57}$Cu$-$$^{57}$Ni situated adjacent to the doubly magic nucleus $^{56}$Ni. 
The reactions were studied in inverse kinematics utilizing a 
beam of radioactive $^{56}$Ni ions in conjunction with the GRETINA 
$\gamma$-array.
Spectroscopic factors are compared with new shell-model calculations using a 
full $pf$ model space with the GPFX1A Hamiltonian for the isospin-conserving 
strong interaction plus Coulomb and charge-dependent Hamiltonians.
These results were used to set new constraints on the 
$^{56}$Ni($p$,$\gamma$)$^{57}$Cu reaction rate for explosive burning conditions 
in x-ray bursts, where $^{56}$Ni represents a key waiting point in the 
astrophysical $rp$-process. 

\end{abstract}

\begin{keyword}
X-ray bursts\sep shell model\sep transfer reactions\sep radioactive beams
\end{keyword}

\end{frontmatter}


Doubly magic nuclei represent special cornerstones in our understanding and 
exploration of nuclear structure (see, {\it e.g.}, 
\cite{2012Natur.486..341H,2018PhRvL.121r2501A}). 
These nuclei, and the nuclei in their immediate vicinity, should be well 
described by very pure shell-model configurations. 
In the case of the doubly magic, self-conjugate nucleus $^{56}$Ni, the major 
shell closure at $N=Z=28$ arises from the spin-orbit splitting of the 
$0f_{5/2}$ and $0f_{7/2}$ orbitals. 
There is evidence that $^{56}$Ni has a relatively soft core with significant 
configuration mixing in the shell structure 
\cite{1998PhRvL..81.1588O,2004PhRvC..70e4319Y,2009PhRvL.103j2501C}. 
It exists on the cusp between stability, having a (terrestrial) half-life of 
6.08~d, and particle instability; the neighboring nucleus, $^{57}$Cu, is only 
proton bound in its ground state. 
This special location means $^{56}$Ni can form a key waiting point in explosive 
astrophysical burning scenarios such as x-ray bursters, impeding the rate of 
flow of material further along the proton drip-line via the 
$^{56}$Ni($p$,$\gamma$)$^{57}$Cu reaction in the $rp$ (rapid proton capture) 
process \cite{1981ApJS...45..389W,1994ApJ...432..326V}. 
$^{56}$Ni is the longest-lived waiting point nucleus and historically was 
originally thought to represent the termination point of the $rp$-process 
\cite{1981ApJS...45..389W}.
A direct measurement of the $^{56}$Ni($p$,$\gamma$) breakout reaction is not 
currently feasible with existing radioactive beam intensities. 
Therefore, to constrain the $^{56}$Ni($p$,$\gamma$)$^{57}$Cu reaction rate used 
in astrophysical models, an indirect approach is mandated. 
This requires a knowledge of the shell structure and properties of states in 
$^{57}$Cu involved in the explosive astrophysical temperature burning range 
from $T\sim0.5$--$2.0$~GK.

The mirror nucleus of $^{57}$Cu, $^{57}$Ni, has a ground-state spin-parity of 
$3/2^{-}$, corresponding to a $1p_{3/2}$ neutron shell-model configuration, and 
$5/2^{-}$ and $1/2^{-}$ excited states corresponding to the occupation 
$0f_{5/2}$ and $1p_{1/2}$ neutron shells, respectively 
\cite{1998NDS....85..415B}. 
A pioneering study of the $^{56}$Ni($d$,$p$)$^{57}$Ni transfer reaction, 
measuring the differential cross section in inverse kinematics, found these 
three lowest-lying states were described with relatively pure single-neutron 
configurations with spectroscopic factors $C^{2}S\sim0.9$, with an approximate 
factor of two uncertainties \cite{1998PhRvL..80..676R}. 
This information was then used to estimate the resonance contributions of the 
two low-lying analog excited states in $^{57}$Cu 
\cite{1998PhRvL..80..676R,1996PhRvC..53..982Z} for the 
$^{56}$Ni($p$,$\gamma$)$^{57}$Cu reaction rate, assuming isospin symmetry.
Subsequently, an attempt was made to measure the single-particle strengths of 
these states directly in $^{57}$Cu using the $^{56}$Ni($^{3}$He,$d$)$^{57}$Cu 
proton transfer reaction; however, due to the limited resolution of 
$\sim\!700$~keV in excitation energy and low statistics, definite conclusions 
could not be drawn \cite{2009PhRvC..80d4613J}. 
Two higher-lying $5/2^{-}$ and $7/2^{-}$ states in $^{57}$Cu are expected to 
dominate the $^{56}$Ni($p$,$\gamma$)$^{57}$Cu reaction rate for burning 
temperatures $T>1$~GK (see, {\it e.g.}, ref. \cite{2001PhRvC..64d5801F}) but 
only limits were set for the experimental $C^{2}S$ values in $^{57}$Ni 
\cite{1998PhRvL..80..676R}.
Calculations predict the combined resonant capture reaction rate on the four 
lowest-lying excited states in $^{57}$Cu is expected to dominate (by 3 orders 
of magnitude) over the direct-capture contribution in explosive astrophysical 
burning conditions \cite{2001PhRvC..64d5801F,1994ApJ...432..326V}.

Here, we present a first study of the $^{56}$Ni($d$,$n$)$^{57}$Cu proton 
transfer reaction aimed at the first direct experimental determinations of the 
proton single-particle strengths of the key low-lying excited states in 
$^{57}$Cu by angle-integrated cross-section measurements. 
The method exploits the high resolution and efficiency of the GRETINA 
(Gamma-Ray Energy Tracking In-beam Nuclear Array) device 
\cite{2013NIMPA.709...44P} and an intense $^{56}$Ni radioactive beam produced 
in flight by the National Superconducting Cyclotron Laboratory (NSCL). 
This new approach has been shown to work successfully in determining 
spectroscopic factors, including key astrophysical resonances 
\cite{2016EPJA...52....6K,2017PhLB..769..549K}, and is described for the 
present application in detail below. 
In addition, we have performed a measurement of the $^{56}$Ni($d$,$p$)$^{57}$Ni 
transfer reaction to study analog states in the mirror nucleus $^{57}$Ni, and 
explore evidence for isospin symmetry breaking effects. 
Furthermore, the $^{56}$Ni($d$,$p$) data were also used to measure for the 
first time the single-particle strength of a high-lying excited state in 
$^{57}$Ni whose analog likely determines the explosive astrophysical burning 
rate above $T\sim1$~GK in the $^{56}$Ni($p$,$\gamma$)$^{57}$Cu reaction. 

The $^{56}$Ni($d$,$n$)$^{57}$Cu and $^{56}$Ni($d$,$p$)$^{57}$Ni reaction 
studies were performed in inverse kinematics.  
A 33.6~MeV/u $^{56}$Ni$^{28+}$ beam was produced by in-flight fragmentation of 
a 28-pnA 160-MeV/u primary beam of $^{58}$Ni$^{27+}$ ions which impinged upon a 
1316-mg/cm$^{2}$ thick $^{9}$Be production target.  
The A1900 fragment separator \cite{2003NIMPB.204...90M} selected the ions of 
interest based on their magnetic rigidity and used a 150-mg/cm$^{2}$ thick Al 
achromatic wedge to provide isotopic separation at its focal plane.  
The resulting $^{56}$Ni$^{28+}$ beam had a purity of 47\% (contaminated mainly 
by $^{55}$Co with traces of $^{52}$Mn and $^{51}$Cr ions) and an average 
intensity of $3\times10^{5}$~$^{56}$Ni particles per second. 
The $^{56}$Ni beam impinged on a 10.7(8)-mg/cm$^{2}$ thick deuterated 
polyethylene target, (CD$_{2}$)$_{n}$, which was surrounded by the GRETINA 
detectors \cite{2013NIMPA.709...44P} positioned in two rings at laboratory 
angles of $58^{\circ}$ and $90^{\circ}$ with respect to the beam direction.  
Beam-like residues were collected and analyzed with the S800 spectrograph 
\cite{2003NIMPB.204..629B} positioned at $0^{\circ}$ scattering angle.  
The analysis line to the S800 was operated in achromatic mode to obtain a total 
acceptance of nearly 100\% for $^{57}$Cu$^{29+}$ and 50(10)\% for 
$^{57}$Ni$^{28+}$ run settings (the lower value for $^{57}$Ni reflects the need 
to block out scattered $^{56}$Ni$^{28+}$ ions from part of the focal plane).  
Using the S800, we measured the intensity of $^{56}$Ni$^{28+,27+,26+}$ species 
after the CD$_{2}$ target relative to the number of incident $^{56}$Ni$^{28+}$ 
beam ions, where we observed that 80\% of the incident ions emerged in the 
fully-stripped charge state.
Given the similar properties of energy, mass, and nuclear charge of the 
$^{56}$Ni beam and heavy residues $^{57}$Cu and $^{57}$Ni, we estimated an 
80(5)\% and 40(11)\% collection and detection efficiency for $^{57}$Cu$^{29+}$ 
and $^{57}$Ni$^{28+}$ ions, respectively. 
To account for reactions on carbon producing the residues of interest, 
measurements were also performed for approximately half the duration of the 
CD$_{2}$ runs with an 8.8(15)-mg/cm$^{2}$ thick (CH$_{2}$)$_{n}$ target.

\begin{figure}
  \subfigure{\includegraphics[angle=0, scale=.47, clip=true, trim=0 0 0 
0]{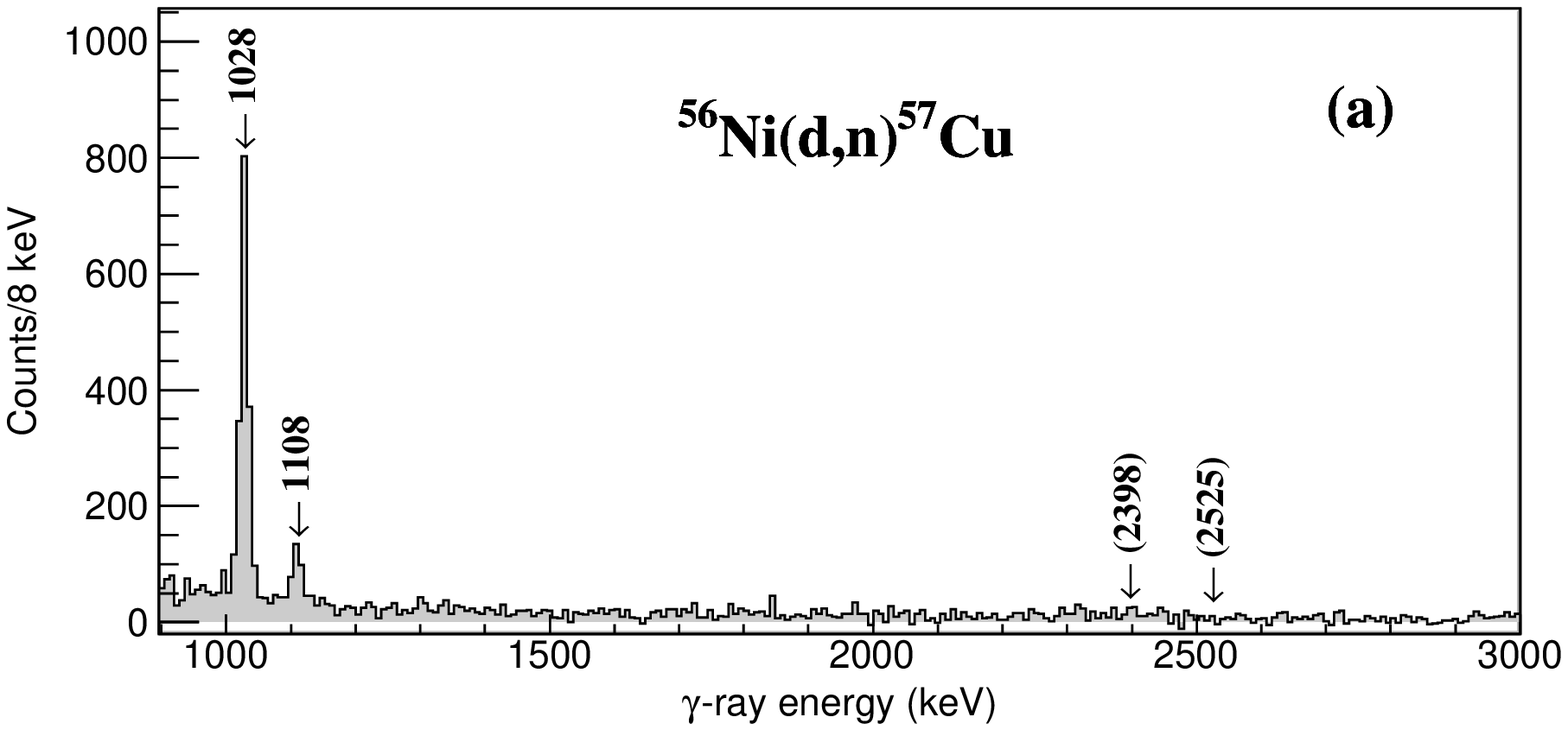}}
  \subfigure{\includegraphics[angle=0, scale=.47, clip=true, trim=0 0 0 
0]{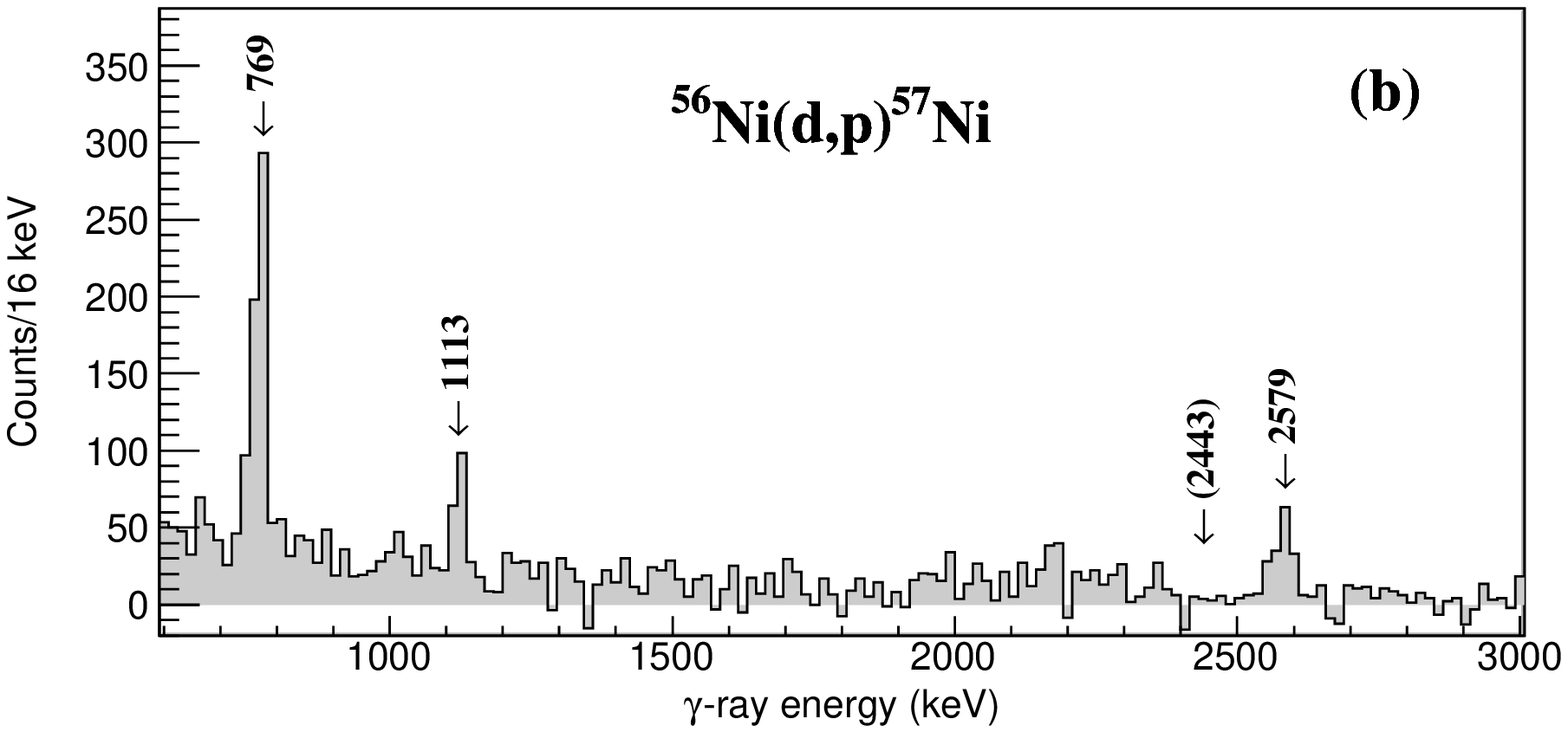} }
\caption[]{
Doppler-reconstructed $\gamma$-ray spectra gated in coincidence with (a) 
$^{57}$Cu and (b) $^{57}$Ni recoils ($v/c\approx 0.25$).
The spectra were produced using the CD$_{2}$ data minus the (scaled) CH$_{2}$ 
data.
The location of known states are indicated.
}
\label{fig:gamma}
\end{figure}
  
The data acquisition system was triggered either by a residue-$\gamma$ 
coincidence or downscaled (by a factor of 5) residue singles events.  
The GRETINA absolute singles efficiency was calibrated with $^{56}$Co and 
$^{152}$Eu sources as described in ref. \cite{2017NIMPA.847..187W}.
An efficiency of 5.5\% for $E_{\gamma}=1332$~keV was achieved for the 
nine-module setup employed here.  
The in-beam GRETINA efficiency was boosted by a factor 1.06(5) compared to a 
stationary source.  
Doppler-adjusted $\gamma$-ray energy spectra gated on $^{57}$Cu and $^{57}$Ni 
ions are shown in Figs. \ref{fig:gamma}a) and b), respectively. 
Events induced by reactions on carbon atoms have already been subtracted, 
following the procedure in refs. 
\cite{2016EPJA...52....6K,2017PhLB..769..549K}. 
In the $^{57}$Cu spectrum, two strong peaks are observed at energies of 1028(1) 
and 1109(2)~keV, which are assigned to the $\gamma$ decays to the ground state 
of the first and second excited states at 1028(4) and 1106(4) keV observed by 
Zhou {\it et al.} in a study of the $^{1}$H($^{58}$Ni,$^{57}$Cu)2$n$ reaction 
\cite{1996PhRvC..53..982Z}. 
There is no evidence for the $\gamma$ decay at 2398(10)~keV reported in the 
same study, or for previously unobserved $\gamma$ decays from the excited state 
at 2525(17)~keV \cite{1987ZPhyA.326..139S,Jokinen2002}.
In Fig.~\ref{fig:gamma}b, three peaks are observed at energies of 768(1), 
1122(5), and 2579(4) keV. 
The first two are assigned to the known decays of the $5/2^{-}$ and $1/2^{-}$ 
levels at 769 and 1113 keV in $^{57}$Ni 
\cite{1998NDS....85..415B,1998PhRvL..80..676R}, respectively. 
The third peak is assigned to the known decay of the $7/2^{-}$ state at 
2578~keV \cite{1998NDS....85..415B}. 
Angle-integrated cross sections for the four lowest excited states in $^{57}$Cu 
and $^{57}$Ni are shown in Table~\ref{tab:sigma} assuming 100\% $\gamma$ 
branches to the ground states (consistent with nuclear compilation data for 
$^{57}$Ni \cite{1998NDS....85..415B}), and insignificant $\gamma$-decay feeding 
from higher-lying excited states (there are no other $\gamma$-ray lines 
observed). 
The overall errors for the cross sections in Table \ref{tab:sigma} were 
obtained from a combination of statistical uncertainties for individual 
transitions, and an uncertainty of 20\% (28\%) for $^{57}$Cu ($^{57}$Ni) 
estimated by combining contributions to the uncertainty from target 
thicknesses, Doppler-corrected $\gamma$-ray detection efficiency, residual 
charge-state distributions, and the momentum acceptance, in quadrature.

  \begin{table}
  \caption{\label{tab:sigma}
  Angle-integrated experimental ($\sigma_{\rm exp}$) and theoretical 
($\sigma_{\rm th}$) reaction cross sections and derived spectroscopic factors 
($C^{2}S_{(d,n)}$, $C^{2}S_{(d,p)}$).  Comparisons are made with shell-model 
(SM) calculations.}
  \footnotesize
  \centering
  \begin{tabular}{lllllll} \hline
  \multicolumn{7}{c}{$^{56}$Ni($d$,\,$n$)$^{57}$Cu} \\ \hline
  $E_{\rm ex}$ & $J^{\pi}$ & $\ell$ & $\sigma_{\rm exp}$ (mb) & $\sigma_{\rm 
th}$ (mb) & $C^{2}S_{(d,n)}$ & $C^{2}S_{\rm SM}$ \\ \hline
  1.028	&  5/2$^{-}$ & 3 & 2.00(40)     & 2.62      & 0.76(28)        & 0.75   
\\
  1.109	&  1/2$^{-}$ & 1 & 0.28(6)     & 0.45      & 0.62(22)        & 0.71  \\
  2.398	&  5/2$^{-}$ & 3 & $<$0.2  & 2.61          & $<8  \times10^{-2}$   & 
$1.8\times10^{-3}$  \\
  2.525	&  7/2$^{-}$ & 3 & $<$0.2  & 14.5         & ---  & $3.9\times10^{-2}$ 
\\    
\hline \hline
  \multicolumn{7}{c}{$^{56}$Ni($d$,\,$p$)$^{57}$Ni} \\ \hline
  $E_{\rm ex}$ & $J^{\pi}$ & $\ell$ & $\sigma_{\rm exp}$ (mb) & $\sigma_{\rm 
th}$ (mb) & $C^{2}S_{(d,p)}$ & $C^{2}S_{\rm SM}$ \\ \hline
  0.768	&  5/2$^{-}$ & 3 & 2.10(60)  & 2.77  & 0.77(31)    & 0.74   \\
  1.122	&  1/2$^{-}$ & 1 & 0.50(15)  & 0.68  & 0.73(31)  & 0.69  \\
  2.443	&  5/2$^{-}$ & 3 & $<$0.4    & 2.61  & $<0.1$          & 
$3\times10^{-4}$  \\
  2.579	&  7/2$^{-}$ & 3 & 1.24(36)  & 14.9  & $8(3)     \times10^{-2}$       & 
$4.1\times10^{-2}$ \\    
\hline 
  \end{tabular}
  \end{table}

The theoretical angle-integrated single-particle ($C^{2}S=1$) cross sections 
for the $^{56}$Ni($d$,$n$)$^{57}$Cu and $^{56}$Ni($d$,$p$)$^{57}$Cu reactions 
shown in Table~\ref{tab:sigma} were calculated using the finite-range adiabatic 
approximation \cite{1974NuPhA.235...56J}, which incorporates deuteron breakup. 
A laboratory beam energy of 32 MeV/u was used corresponding to the approximate 
center-of-target energy. 
Nucleon-target interactions used the CH89 \cite{1991PhR...201...57V} optical 
potential and the nucleon-nucleon interaction from ref. 
\cite{1968AnPhy..50..411R}.
The $n(p)\,+^{56}$Ni final states were described by a real Woods-Saxon 
potential with central and spin-orbit terms. 
The radius and diffuseness of these potentials were set to 1.23~fm and 0.67~fm, 
respectively. 
The spin-orbit term was given the standard depth of $V_{\rm so} = 6$~MeV.
For the ($d$,$p$) calculations, the central potential depth was adjusted to 
reproduce the final bound-state binding energies. 
For the ($d$,$n$) calculations, the final states are resonances. 
Here, we applied a bound-state approximation in which we adjusted the depth of 
the central potential to produce a final state bound by just $E=0.001$~MeV, as 
was done in \cite{2016EPJA...52....6K}. 
For low-lying resonances, this approximation introduces less than 1\% error, 
but for the two higher-lying resonances, this approximation can introduce an 
error of $\sim\!6$\%. 
Repeating the calculation with the Becchetti-Greenlees potential 
\cite{1969PhRv..182.1190B} changed the total cross section by about 15\%, 
dominating the error introduced by the bound-state approximation. 
Based on this and other studies \cite{2011PhRvC..83c4610N}, we estimate an 
error of up to 30\% in the total cross section calculations. 
The effective adiabatic potentials for ($d$,$p$) and ($d$,$n$) were computed 
with {\sc twofnr} \cite{TWOFNR} and the transfer calculations were performed 
with the reaction code, {\sc fresco} \cite{1988CoPhR...7..167T}.
Shell-model wavefunctions for $^{56,57}$Ni and $^{57}$Cu were obtained in the 
full $  pf  $ model space with the GPFX1A Hamiltonian 
\cite{2005EPJAS..25..499H} for the isospin-conserving strong interaction plus 
the Coulomb and charge-dependent Hamiltonians from ref. 
\cite{1989NuPhA.491....1O}.
The spectroscopic factors (shown in Table~\ref{tab:sigma}) were derived from 
the 
overlap of these wavefunctions.

  \begin{table}
  \caption{\label{tab:astro}
  Resonance parameters used in the $^{56}$Ni($p$,\,$\gamma$)$^{57}$Cu reaction 
rate calculation.  See the text for details.
  }
  \footnotesize
  \centering
  \begin{tabular}{llllll} \hline 
  $E_{\rm ex}$ (keV) & $E_{\rm r}$ (keV)  & $J^{\pi}$ & $\Gamma_{p}$ (eV) & 
$\Gamma_{\gamma}$ (eV) & $\omega\gamma$ (eV)\\ \hline
  1028(1)	& 338	& 5/2$^{-}$  & $5.7\times10^{-12}$ & $1.9\times10^{-4}$ 
& $1.7\times10^{-11}$  \\
  1108(2)	& 418	& 1/2$^{-}$  & $1.9\times10^{-7}$  & $8.6\times10^{-3}$ 
& $1.9\times10^{-7}$  \\
  2398(10)	& 1708	& 5/2$^{-}$  & $5.5\times10^{-3}$  & $9.0\times10^{-3}$ 
& $1.0\times10^{-2}$  \\
  2525(17)	& 1835	& 7/2$^{-}$  & $5.3\times10^{-1}$  & $6.8\times10^{-3}$ 
& $2.7\times10^{-2}$  \\    
\hline 
  \end{tabular}
  \end{table}

Experimental spectroscopic factors for the two lowest-lying excited states in 
$^{57}$Cu are reported here for the first time. 
The values agree very well with shell-model calculations and are consistent 
with strong single particle states. 
A level hierarchy with the $5/2^{-}$ level below the $1/2^{-}$ is confirmed by 
the relative difference in cross sections for these states, which are 
consistent with our reaction theory calculations. 
Previously this hierarchy had been assumed largely on the basis of mirror 
energy level shift arguments \cite{1996PhRvC..53..982Z,2001PhRvC..64d5801F}. 
This relative difference in cross section is also reflected for the analog 
states in $^{57}$Ni. 
The spectroscopic factors for these states in $^{57}$Ni are found to be very 
similar to the analog states in $^{57}$Cu, and show no evidence of significant 
isospin symmetry breaking. 
The $^{57}$Ni $C^{2}S$ values are lower, and more precise, than those first 
reported by Rehm {\it et al.} \cite{1998PhRvL..80..676R}, but agree well within 
errors. 
A clear difference between the spectra in Fig.~\ref{fig:gamma} is the strong 
presence of the decay of the $7/2^{-}$ state in $^{57}$Ni which is not present 
for $^{57}$Cu. 
From this we can infer that proton decay represents the dominant branch of the 
$7/2^{-}$ state in $^{57}$Cu (we therefore do not present a $C^{2}S$ value for 
this state in $^{57}$Cu in Table \ref{tab:sigma}). 
The $C^{2}S(\ell=3)$ value for the $7/2^{-}$ state in $^{57}$Ni represents a 
first measurement (rather than an upper limit) and is an interesting test of 
the shell-model calculations. 
It is found to be much weaker in single particle strength than the lower-lying 
states and is broadly consistent with the shell-model prediction.
The shell-model calculations suggest the neighboring $5/2^{-}$ state has a 
smaller single-particle strength consistent with its non-observation here in 
both the $^{57}$Ni and $^{57}$Cu data. 
Interestingly, the $\gamma$ decay of this state was clearly observed in the 
$^{1}$H($^{58}$Ni,$^{57}$Cu)2$n$ reaction \cite{1996PhRvC..53..982Z}, which 
suggests the $\gamma$ branch for this state in $^{57}$Cu is at least comparable 
to the proton branch, otherwise it would not have been seen in that study. 

Table~\ref{tab:astro} shows resonance parameters for the four lowest-lying 
excited states in $^{57}$Cu used to calculate the 
$^{56}$Ni($p$,$\gamma$)$^{57}$Cu astrophysical reaction rate. 
Shell-model calculations of the $\gamma$-decay widths used the M1 and E2 
effective operators from ref.~\cite{2004PhRvC..69c4335H}.
The uncertainty in the $\gamma$-decay widths is about a factor of two 
\cite{2004PhRvC..69c4335H}.
Proton-decay widths were calculated from $  \Gamma  = C^{2}S \cdot P \cdot 2W  
$,
where 
$  P  $ are the penetration factors
obtained from the Coulomb wavefunction, and $  W  $ is the Wigner
single-particle width. The $\ell$-dependent radius for the evaluation of $P$ 
was chosen to reproduce the single-particle width obtained for proton 
scattering from a Woods-Saxon potential.
The Woods-Saxon parameters were consistent with the geometry used in the cross 
section calculations and were chosen to reproduce the proton separation energy 
of $^{56}$Ni to $^{55}$Co and the rms charge radius of $^{58}$Ni.
The potential depth for the scattered proton was adjusted to give a resonance 
$Q$-value of 1~MeV.
The uncertainty in these calculations of the single-particle proton-decay 
widths is $\sim\!20$\%.
A proton separation energy of 690.3(4)~keV \cite{2017ChPhC..41c0003W} was used 
for $^{57}$Cu to calculate the resonance energies, which incorporate the new 
more precise excitation energy values reported here for the two lowest-lying 
excited states. 
The proton widths for these states were calculated for the first time using the 
experimentally constrained $C^{2}S$ values obtained for $^{57}$Cu in the 
present work. 
For the $7/2^{-}$ state we derive the proton width value using the 
$C^{2}S(\ell=3)$ value reported here for the first time for the analog state in 
$^{57}$Ni. 
For the proton width of the 2398-keV $5/2^{-}$ level, we take the $C^{2}S$ 
value from the shell-model calculation which is compatible with the 
experimental observational limit in the present $^{57}$Ni data (see 
Table~\ref{tab:sigma}). 

\begin{figure}
  \centering
  \includegraphics[angle=0, scale=0.47, clip=true, trim=0 0 0 30]{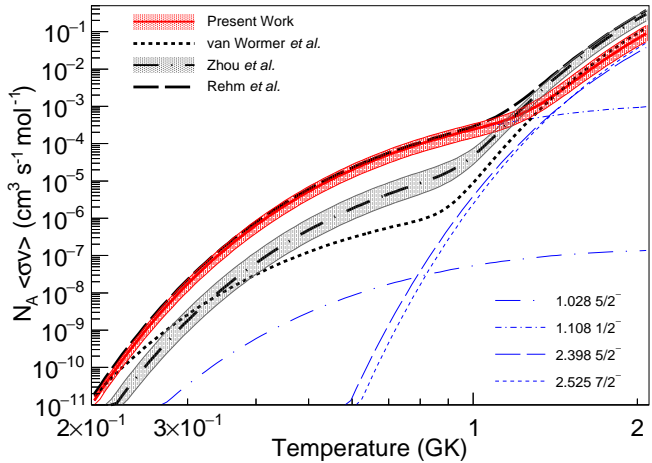}
\caption[]{
$^{56}$Ni($p$,$\gamma$) astrophysical reaction rate and uncertainty bounds 
calculated using the resonance parameters listed in Table \ref{tab:astro} 
(shown in orange).  The total rates from refs. 
\cite{1994ApJ...432..326V,1996PhRvC..53..982Z,1998PhRvL..80..676R} are shown 
for comparison; additionally, as the uncertainty bounds were presented in ref. 
\cite{1996PhRvC..53..982Z}, these are also depicted.  See text for details.
}
\label{fig:rate}
\end{figure}

The present $^{56}$Ni($p$,$\gamma$)$^{57}$Cu reaction rate and uncertainty 
bounds incorporating our new experimental results and shell-model calculations 
is shown in Fig.~\ref{fig:rate}. 
Our rate for $T<1$~GK is dominated by the $\ell=1$ capture resonance at 
417.7(1.8) keV for which we derive a resonance strength, 
$\omega\gamma=1.9(8)\times10^{-7}$~eV. 
For the region with $T>1$~GK, both the $\ell=3$ captures on the higher-lying 
$5/2^{-}$ and $7/2^{-}$ states are found to contribute to, and collectively 
dominate, the reaction rate. 
The $5/2^{-}$ state is the only lower-lying resonance for which we have only an 
experimental upper limit on the spectroscopic factor. 
If this value is significantly lower than the shell-model calculation this 
would reduce its strength/significance relative to the $7/2^{-}$ state. 
A much higher value for $C^{2}S$, and therefore the proton partial width, is 
deemed unlikely as $\gamma$ decay has been observed from this $5/2^{-}$ state 
in $^{57}$Cu \cite{1996PhRvC..53..982Z}.
Figure \ref{fig:rate} also shows some previous reaction rate calculations for 
comparison \cite{1994ApJ...432..326V,1996PhRvC..53..982Z,1998PhRvL..80..676R}. 
At lower temperatures, the van Wormer theoretical calculation is orders of 
magnitude lower and reflects the then unknown resonance energies in $^{57}$Cu. 
The later rate of Zhou {\it et al.} used known experimental energies but the 
proton and $\gamma$ widths were entirely estimated from theory; the large 
reaction rate difference, particularly for $T<1$~GK, is caused by the 
relatively low proton partial width derived for the $1/2^{-}$ resonance in that 
estimate \cite{1996PhRvC..53..982Z}. 
The rate of Rehm {\it et al.} \cite{1998PhRvL..80..676R} incorporates the 
results for the spectroscopic factors for the two lowest-lying excited states 
in $^{57}$Ni. 
This central rate agrees well at low temperatures, as the same proton width and 
resonance strength values are obtained for the $1/2^{-}$ state. 
One should note that the spectroscopic factor used for the calculation in ref. 
\cite{1998PhRvL..80..676R} was based on the analog state in $^{57}$Ni, and had 
a higher $C^{2}S$ value (and uncertainty) than that measured directly in the 
present study on $^{57}$Cu. 
For the two higher-lying resonance contributions refs. 
\cite{1998PhRvL..80..676R} and \cite{1996PhRvC..53..982Z} adopted the same 
partial width values so their total rates for $T>1$~GK are nearly the same. 
Our result is significantly lower in this region, reflecting the lower partial 
widths obtained (particularly for the $5/2^{-}$ level) in the present 
shell-model calculations. 

In summary, we report angle-integrated cross-section measurements for the 
$^{56}$Ni($d$,$n$)$^{57}$Cu and $^{56}$Ni($d$,$p$)$^{57}$Ni transfer reactions. 
Comparisons with reaction theory calculations allow definitive assignments of 
the first two excited states in $^{57}$Cu to their analogs in the mirror 
partner, $^{57}$Ni. 
First measurements of the spectroscopic factors for these two states in 
$^{57}$Cu show they have a strong single-particle character with values 
agreeing well with the new shell-model calculations obtained using a full $pf$ 
model space with the GPFX1A interaction plus Coulomb and charge-dependent 
Hamiltonians.
From a comparison with their analog states in $^{57}$Ni we find no evidence for 
significant isospin symmetry breaking effects.
The spectroscopic factor of a high-lying $7/2^{-}$ state is determined for the 
first time in $^{57}$Ni and is found to have a much weaker single particle 
character in reasonable agreement with the new shell-model calculations.
We use these new results to re-evaluate and significantly constrain the 
$^{56}$Ni($p$,$\gamma$)$^{57}$Cu astrophysical reaction rate required for 
modeling of explosive burning in x-ray bursts where the astrophysical 
$rp$-process is thought to occur.

\section*{Acknowledgments}
This work was supported by the U.S. National Science Foundation (NSF) under 
Cooperative Agreement No. PHY-1102511 (NSCL) and No. PHY-1430152 (JINA-CEE). 
GRETINA was funded by the DOE, Office of Science. 
Operation of the array at NSCL was supported by the DOE under Grants No. 
DE-SC0014537 (NSCL) and No. DE-AC02-05CH11231 (LBNL).
BAB kindly acknowledges funding from an NSF Grant No. PHY-1811855.
The Edinburgh group is appreciative of funding from the UK STFC 
under Grant No. ST/L005824/1.

\end{document}